\documentclass[aps,preprintnumbers,amsmath,amssymb,showpacs,nofootinbib,twocolumn]{revtex4}
\usepackage{graphicx}
\usepackage{dcolumn}
\usepackage{bm}

\begin{document}
\title{Universal behaviour of the \boldmath $\gamma^*\gamma\to (\pi^0,\eta,\eta')$ transition form factors}
\author{Dmitri Melikhov$^{1,2,3}$ and Berthold Stech$^4$}
\affiliation{
$^1$HEPHY, Austrian Academy of Sciences, Nikolsdorfergasse 18, A-1050  
Vienna, Austria\\
$^2$Faculty of Physics, University of Vienna, Boltzmanngasse 5, A-1090  
Vienna, Austria\\
$^3$SINP, Moscow State University, 119991 Moscow, Russia\\
$^4$ ITP, Heidelberg University, Philosophenweg 16, D-69120,  
Heidelberg, Germany}
\date{\today}

\begin{abstract}
The photon transition form factors of $\pi$, $\eta$ and $\eta'$ are discussed in view of recent measurements. 
It is shown that the exact axial anomaly sum rule allows a precise comparison of all three form 
factors at high-$Q^2$ independent of the different structures and distribution amplitudes of the 
participating pseudoscalar mesons. We conclude: 
(i) The $\pi\gamma$ form factor reported by Belle is in excellent agreement with the non-strange $I=0$ 
component of the $\eta$ and $\eta'$ form factors obtained from the BaBar measurements.
(ii) Within errors, the $\pi \gamma$ form factor from Belle is compatible with the asymptotic pQCD behavior, 
similar to the $\eta$ and $\eta'$ form factors from BaBar. Still, the best fits to the data sets of $\pi\gamma$, 
$\eta\gamma$, and $\eta'\gamma$ form factors favor a universal small logarithmic rise $Q^2 F_{P\gamma}(Q^2)\sim \log(Q^2)$.
\end{abstract}
\maketitle

\section{Introduction and results}
\vspace{-0.25cm}
The extensive experimental study of the $\gamma^*\gamma\rightarrow P$ reactions ($P=\pi^0,\eta,\eta'$) 
\cite{cello,cleo,babar2,babar,babar1} attracted much attention from theorists 
(for recent references see \cite{radyushkin,roberts,dorokhov,agaev,teryaev2,bt,kroll,mikhailov,blm,lcqm,czyz,ms2012}). 
The reason was the report of the BaBar collaboration \cite{babar} about a persistent rise
of the combination $Q^2F_{\pi\gamma}(Q^2)$ in the $Q^2$ region from $10$ GeV$^2$ to $40$ GeV$^2$. The measured form factor surpassed 
the asymptotic behaviour $Q^2F_{\pi\gamma}(Q^2)\to \sqrt{2}f_\pi$ \cite{bl}, $f_\pi= 0.131$ GeV, predicted by perturbative 
QCD (pQCD). Several theoretical investigations \cite{roberts,bt,mikhailov,blm} indicated, however,
that the corresponding increase  of the $\pi\gamma$ form factor for large $Q^2$ values is hard to explain. 
Very recently new experimental information came from the Belle collaboration \cite{belle2012}. 
The data for  $Q^2F_{\pi\gamma}(Q^2)$ presented by this group show only a 
very mild (if any) increase in the high-$Q^2$ region.

In this situation a comparison of  the $\pi\gamma$ form factor with the ones for $\eta$ and $\eta'$ can be helpful. At first sight,  
this appears difficult because of the different structure (different quark distribution amplitudes) of these particles. 
In the present note we show however, that by using the exact anomaly sum rule, the high-$Q^2$ behavior of $\pi$, 
$\eta$ and $\eta'$ form factors is essentially determined by the well-known lowest order 
spectral representation of the triangle quark diagram. Therefore, a comparison of the three form factors is possible 
and can be trusted.

The application of the anomaly sum rule requires to relate the full-QCD spectral densities with the spectral densities 
obtained from perturbative QCD. This can be achieved by using the concept of duality. This way the anomaly sum rule offers the 
interesting possibility \cite{teryaev2} to calculate the transition form factors without referring to the QCD factorization 
theorem. No assumptions are needed about the light-cone distribution amplitudes of pseudoscalar mesons with their 
specific end-point behavior and Gegenbauer coefficients. 

The high-$Q^2$ behavior is determined by the high-energy dependence of the spectral density in the corresponding 
integrals, see Eqs.~(\ref{Ct}) and (\ref{FF}) below. At high energy---above the resonance region---the spectral density 
can  be  very well approximated by perturbation theory. This spectral density is the same for the 
three form factors. (The effect of the difference between the current masses of strange and non strange quarks  should  be negligible  at high energy). Thus
we can conclude that at high-$Q^2$ the functional dependence of these form factors should be the same. 
A detailed analysis in \cite{blm} suggests that the universality may be expected already at $Q^2\ge$ 10 GeV$^2$.
Different decay constants provide for different multiplication factors but do not affect the slope. 

\begin{figure}[ht]
\begin{center}
\begin{tabular}{c}
\includegraphics[width=8.1cm]{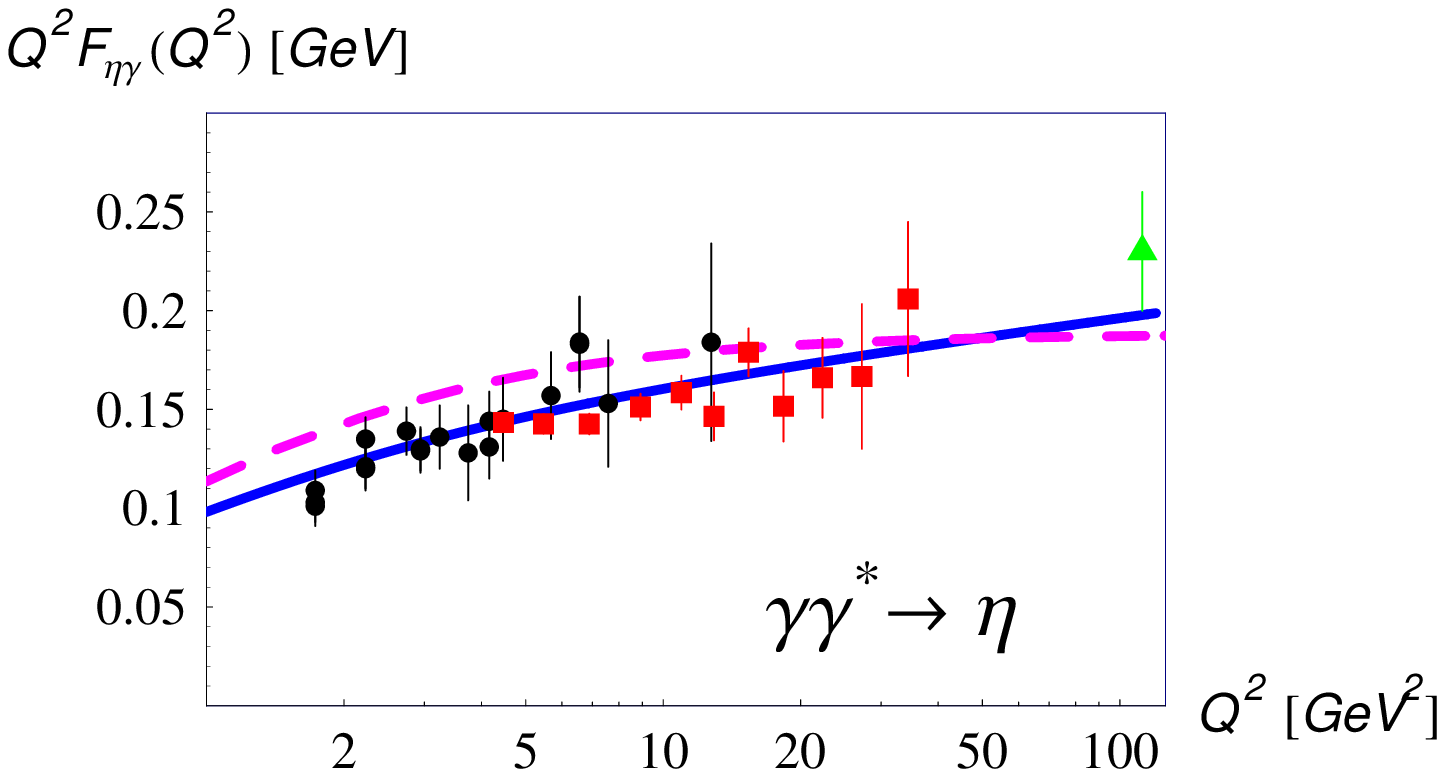} \\
\includegraphics[width=8.1cm]{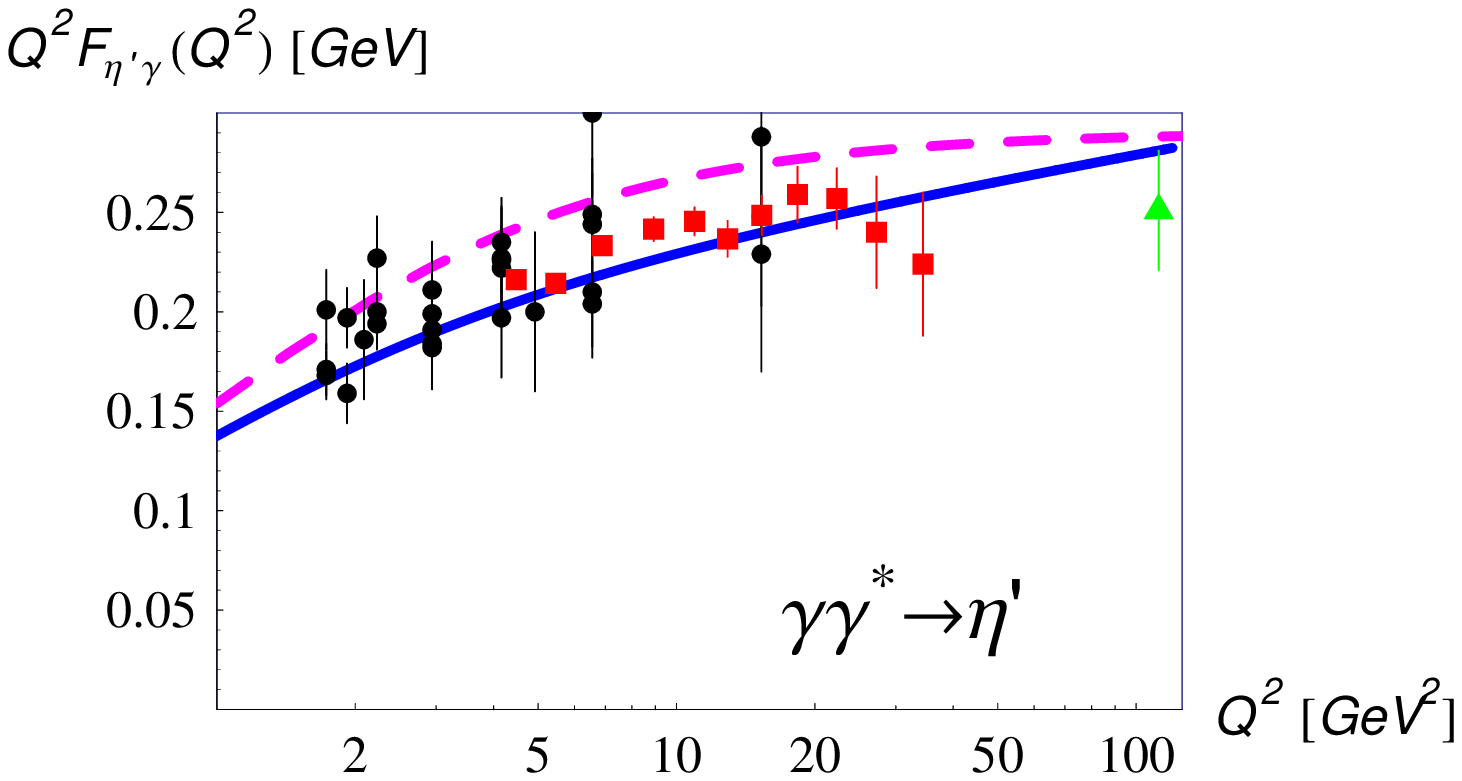}
\end{tabular}
\caption{\label{Plot:1}
Form factors $Q^2 F_{P\gamma}$ ($P=\eta,\eta'$) vs $Q^2$: experimental data  
from Cello and Cleo \cite{cello,cleo} (black dots), BaBar \cite{babar1} (red squares), 
and the data borrowed from the time-like region \cite{babar2} (green triangles). 
Dashed lines - the results from \cite{blm} which obey the factorization theorem at $Q^2\to\infty$; 
solid lines - our fits for $r_q^{(I=0)}=r_s = 0.05 $ GeV$^2$.} 
\end{center}
\end{figure}

Looking now at the data, the following observations can be made:

(i) The BaBar measurements of the $\eta\gamma$ and $\eta'\gamma$ form factors (Fig.~\ref{Plot:1}) are 
within errors compatible with pQCD factorization which implies saturation of the combination $Q^2 F_{P\gamma}(Q^2)$. 
Still, the data seem to indicate a very mild (e.g. logarithmic) rise with $Q^2$. 

\begin{figure}[ht]
\begin{center}
\begin{tabular}{c}
\includegraphics[width=8.2cm]{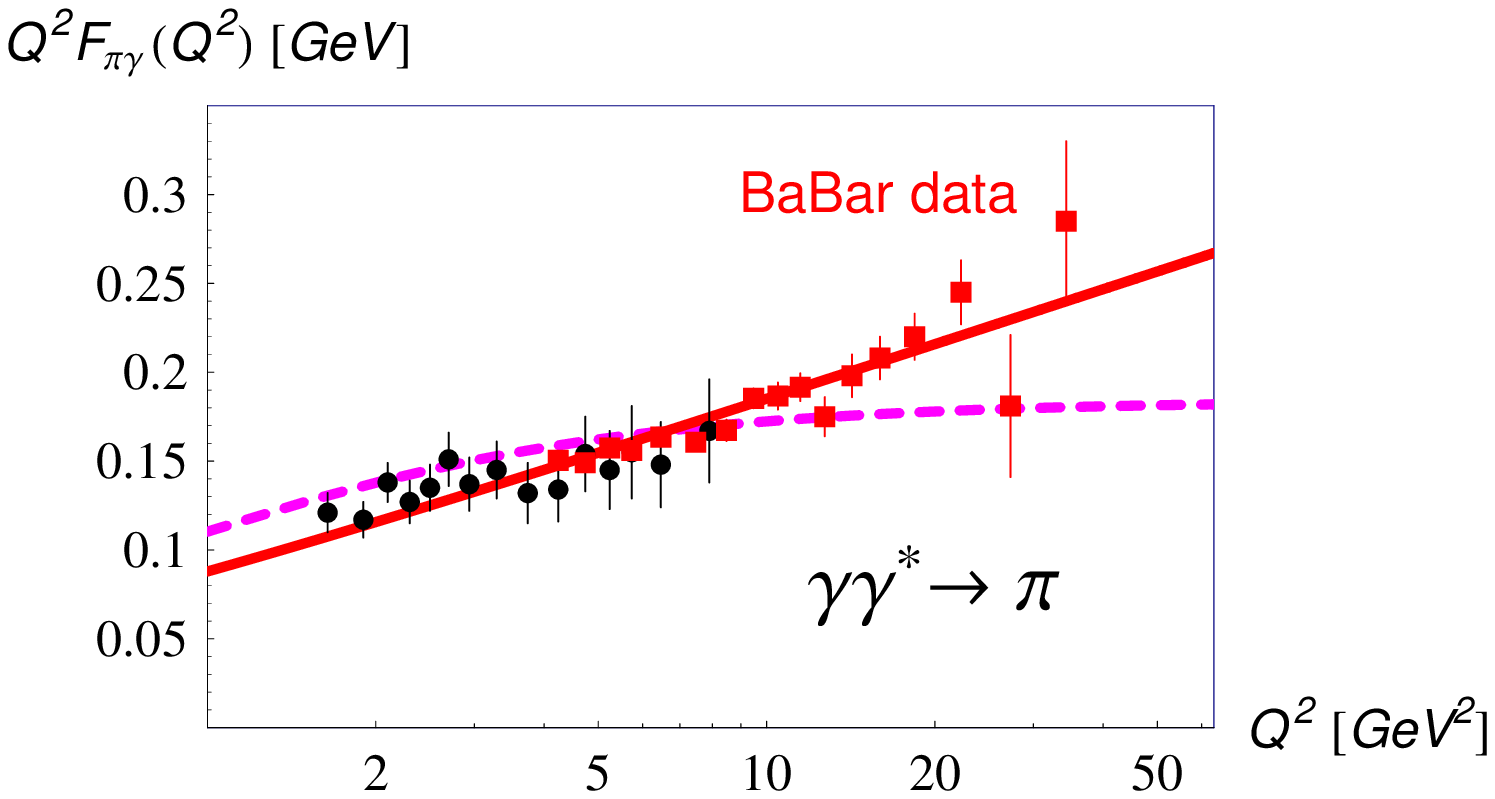}\\ 
\includegraphics[width=8.2cm]{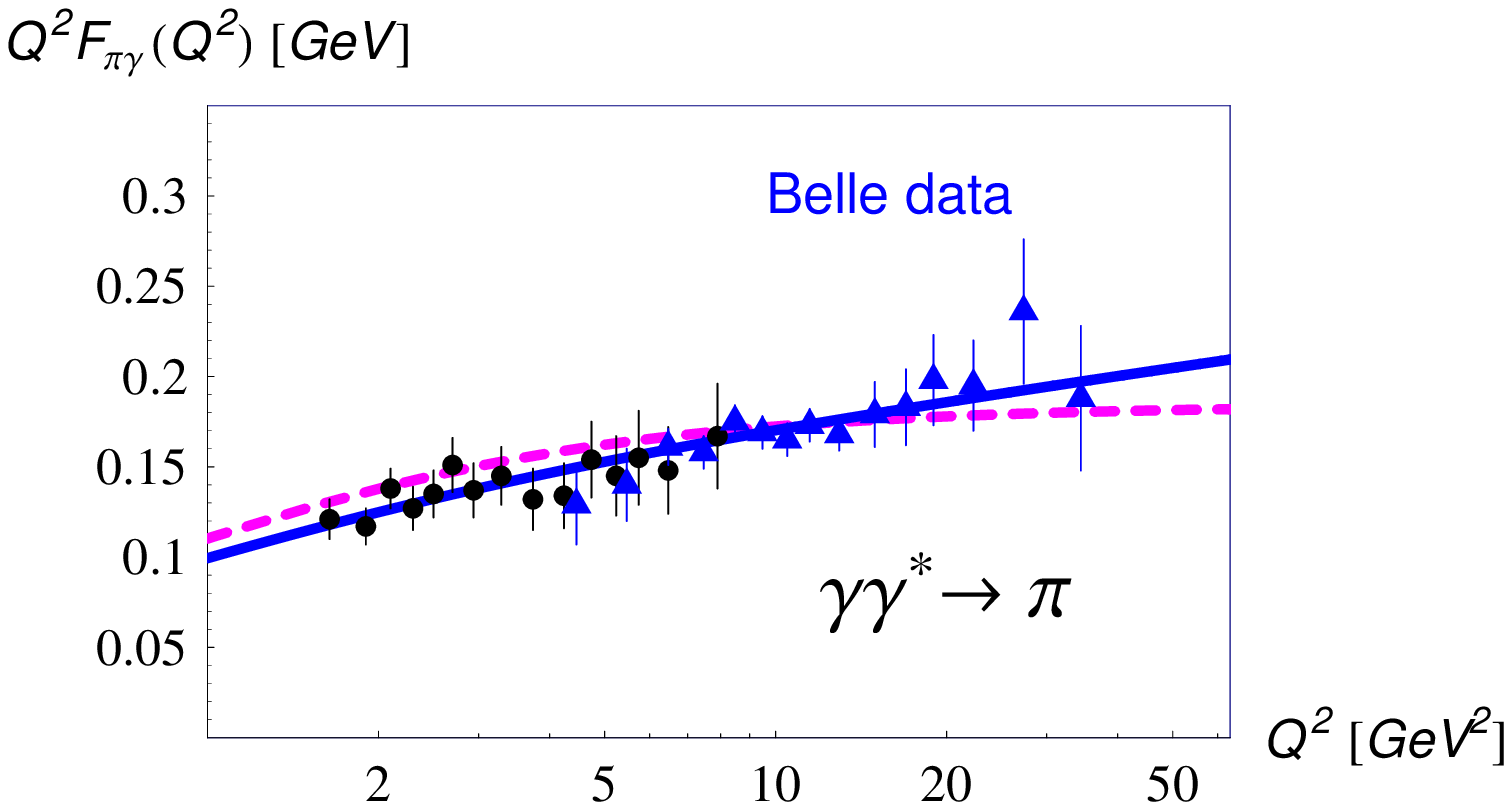}\\
\includegraphics[width=8.2cm]{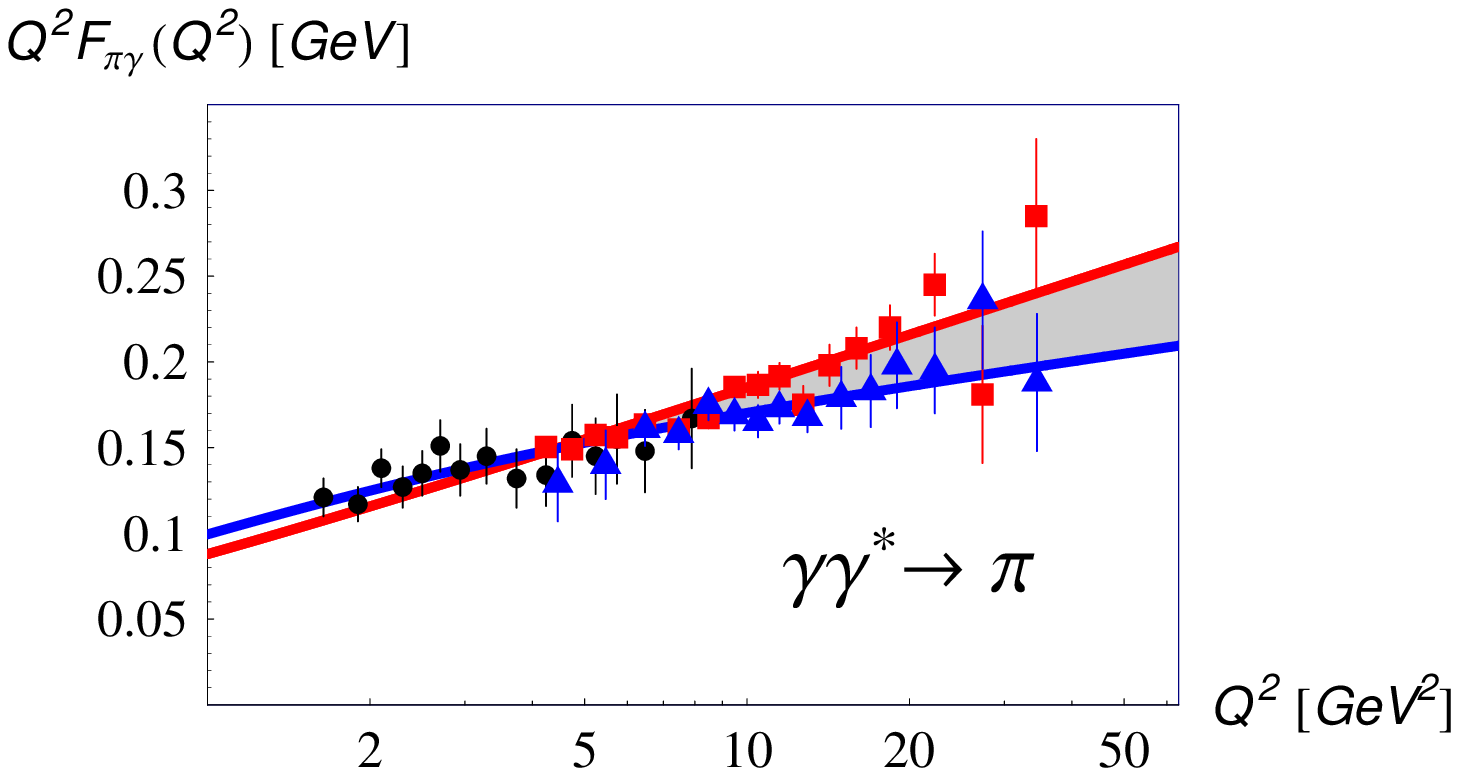}
\end{tabular}
\caption{\label{Plot:0}
Form factor $Q^2 F_{\pi\gamma}(Q^2)$ vs $Q^2$: experimental data from Cleo \cite{cleo} (black dots),  
BaBar \cite{babar} (red squares) and Belle \cite{belle2012} (blue triangles);
dashed line -- the results from \cite{blm} which obey the factorization theorem at $Q^2\to\infty$. 
Solid lines -- our fits. In the fit to the Belle data, the value of $r_q^{(I=1)}$ is taken identical to $r_q^{(I=0)}$ 
for the nonstrange component of $F_{\eta\gamma}$ and $F_{\eta'\gamma}$ in Fig.~\ref{Plot:1}.
The grey shaded region corresponds to the range 0.05 GeV$^2\le r_q^{(I=1)} \le$ 0.17 GeV$^2$.}
\end{center}
\end{figure}

(ii) The large-$Q^2$ behaviour of the form factor $F_{\pi\gamma}(Q^2)$ as observed by 
BaBar (Fig.~\ref{Plot:0}) is in 
some conflict with the saturation predicted by QCD factorization. 
These data  suggest an increase of $Q^2 F_{\pi\gamma}(Q^2)$. The rise as seen from the high-$Q^2$ points 
of the BaBar data is much larger than the rise observed for the $\eta$ and $\eta'$ form factors. 

(iii) The Belle \cite{belle2012} measurement of the $\pi \gamma$ form factor Fig.~\ref{Plot:0} is within 
errors compatible with pQCD factorization which implies saturation of the combination 
$Q^2 F_{P\gamma}(Q^2)$. Still, also these data seem to indicate a logarithmic rise with $Q^2$, 
but a very mild one like in the $\eta$ and $\eta'$ data.

We now apply our theoretical argument which states that the high-$Q^2$ behavior of the three form factors 
should be the same. From the observations (i), (ii) (iii) one can already conclude that this property is 
indeed seen in the data if for $F_{\pi \gamma}$ the Belle data points are used. 
However, for a more quantitative analyses, the slight logarithmic rise indicated by  the data should be 
taken into account.  

To describe the logarithmic rise, the simplest use of the duality 
concept---a replacement of the absorptive part of the form factor beyond the resonance region  
by the lowest-order perturbative QCD spectral density---cannot be maintained: 
To account in a phenomenological way for higher-order effects, 
the lowest-order perturbative spectral density will be be multiplied by a 
correction factor $R(s) $ which goes to one for high values of  $s$.
Here $s$ denotes the square of the energy variable. 
It turns out that it is sufficient to describe $R(s)$ as a simple 
function of a single fit parameter $r$: $R(s) = 1 - \frac{r}{s}$. 
The function $R(s)$ starts at the effective threshold relevant for each process.
(For technical details see the next Section).\footnote{In principle,  
$r$ could be $Q^2$-dependent, e.g., $r(Q^2)=\frac {r_0}{1+ a Q^2}$. 
This choice would restore saturation at large $Q^2$.}

Our simple model for the full set of the form factors thus involves six parameters: 
the three effective thresholds $s_q^{(I=0,1)}$ , $s_s$ for the $I=0$ , $I=1$ and $\bar ss$ 
components of the form factors  and the three $r$ parameters  $r_q^{(I=0,1)}$, $ r_s$.
The decay constants $f_s$ and $f_q$ and the $\eta - \eta' $mixing angle $\phi$ 
are taken from Ref. \cite{feldmann}. 

\vspace{.4cm}
\centerline{\bf\boldmath The form factors of $\eta$ and $\eta'$}
\vspace{.2cm}

Let us first set $r_s=r_q^{(I=0)}=0$, as implied by the pQCD factorization theorem. 
The fit to the existing data yields $s_q^{(I=1)}=0.67\pm 0.07$ GeV$^2$ and $s_s=1.0\pm 0.03$ 
GeV$^2$ with $\chi^2/DOF=72/71$.  
Strictly speaking, in view of the experimental data, further improvements are unnecessary. 
Nevertheless, having in mind the BaBar result for the pion form factor, 
let us allow also nonzero values of $r$ setting $r_q^{(I=0)}=r_s$ 
(at the present accuracy of the data it makes no sense to treat them independently). Then the fitting 
procedure gives $s_q^{(I=0)}=0.57\pm 0.07$ GeV$^2$, $s_s=0.94\pm 0.03$ GeV$^2$ and   
$r_s=r_q^{(I=0)}=0.05\pm 0.01$ GeV$^2$ with $\chi^2/DOF=60/70$.
The corresponding results are shown in Fig.~\ref{Plot:1}. Notice that the threshold 
values obtained by the fits are not far from 
 $s_q^{(I=1)}=0.56$ GeV$^2$ and $s_s=0.76$ GeV$^2$ 
suggested from the relevant elements of the $\eta - \eta' $ 
mass matrix \cite{feldmann}. 


\vspace{.6cm}
\centerline{\bf\boldmath The $\pi^0$ form factor}
\vspace{.2cm}

A fit to the data set containing the Cleo and the BaBar results 
in Fig.~\ref{Plot:0} leads to a much larger paramater 
$r^{(I=1)}=0.14\pm 0.014$ GeV$^2$ than obtained for $r_q^{(l=0)} $ found above. 
The $\chi$ value is $\chi^2/DOF=26/30$ and the threshold parameter $s_q^{(I=1)}=0.32\pm 0.07$ GeV$^2$.
Most important, by setting $r^{(I=1)}=0$ one gets an extremely bad fit with $\chi^2/DOF=88/31$. 
Obviously, the BaBar high-$Q^2$ data are not compatible with the asymptotic pQCD result 
at large $Q^2$ (15-35 GeV$^2$).

In contrast, a fit to the data set containing the Cleo and the Belle data in Fig.~\ref{Plot:0} 
is fully compatible with $r_q^{(I=1)}=0$: for $s_q^{(I=1)}=0.64\pm 0.01$ GeV$^2$, one gets an 
excellent fit to the data with $\chi^2=16/29$. No further improvements are necessary. 
Nevertheless, allowing in addition for a nonzero parameter $r_q^{(I=1)}$ leads to  
$r_q^{(I=1)}=0.06\pm 0.02$ GeV$^2$ and $s_q^{(I=1)}=0.5\pm 0.06$ GeV$^2$ with $\chi^2/DOF=10/28$. 
Remarkably, this value of $r_q^{(I=1)}$ is equal to $r_q^{(I=0)}$ as obtained
from the $\eta$ and $\eta'$ data. Also the effective thresholds for the nonstrange quark sector 
are very close to each other: $s_q^{(I=1)} \simeq s_q^{(I=0)}$.

The shaded region in Fig.~\ref{Plot:0} corresponds to the variation of 
$r$ in the range 0.05 GeV$^2\le r \le$ 0.17 GeV$^2$. 

To summarize, using the Belle data for the large-$Q^2$ region, the previously puzzling difference between 
the $(\eta,\eta')\gamma$ and $\pi\gamma$ form factors is no more present: all three processes---after 
taking particle mixing into account---can be well described by only two effective thresholds and a 
small universal parameter $r$. 

Certainly, more precise measurements are needed to establish a small logarithmic increase of $Q^2 F_{P\gamma}(Q^2)$ 
for high $Q^2$ values indicated by the data and parametrized by a nonzero $r\simeq 0.05$ GeV$^2$. 


\section{Technical details}
We now provide some details of our calculation of the $P\gamma$ form factors. For subtleties, we refer to \cite{ms2012}.

Our starting point is the amplitude
\begin{eqnarray}
\langle 0| j_{\mu}^5|\gamma(q_2)\gamma^*(q_1)\rangle =e^2T_{\mu\alpha\beta}(p|q_1,q_2)
\varepsilon^\alpha_1\varepsilon^\beta_2, \nonumber\\
\qquad p=q_1+q_2.
\end{eqnarray}
Here $\varepsilon_{1,2}$ denote the photon polarization vectors.
This amplitude is considered for $q_1^2 = - Q^2$ and $q_2^2=0 $. 
Its general decomposition contains four independent Lorentz structures, 
but for our purpose only one structure is needed \cite{blm}
\begin{eqnarray}
\label{F}
T_{\mu\alpha\beta}(p|q_1,q_2)= p_\mu \epsilon_{\alpha\beta q_1 q_2} iF(p^2,Q^2) +\dots
\end{eqnarray}
The invariant amplitude $F(p^2,Q^2)$ satisfies the spectral representations in $p^2$ at fixed $Q^2$:
\begin{eqnarray}
F(p^2,Q^2)=\frac{1}{\pi}\int\limits_{s_{\rm th}}^\infty\frac{ds}{s-p^2}\,\Delta(s,Q^2), 
\end{eqnarray}
where $\Delta(s,Q^2)$ is the physical spectral density and $s_{\rm th}$ denotes the physical threshold. 
Perturbation theory yields the spectral density as an expansion in powers of $\alpha_s$.
The lowest order contribution, $\Delta^{(0)}_{\rm pQCD}(s,Q^2)$, corresponds to the one-loop 
triangle diagram with the axial current and two vector currents in the vertices \cite{teryaev,ms,m}
\begin{eqnarray}
\label{1loop}
\Delta^{(0)}_{\rm pQCD}&=& \frac{1}{2\pi}\frac{1}{(s+Q^2)^2}
\left[Q^2\,w+2m^2\log\left(\frac{1+w}{1-w}\right)\right], 
\nonumber\\&&
\quad w=\sqrt{1- 4 m^2/s}.
\end{eqnarray}
Here $m$ denotes the mass of the quark propagating in the loop. 
The integral of $\Delta^{(0)}_{\rm pQCD}(s,Q^2)$ from $s=4m^2$ to infinity is independent of $m$ and $Q^2$ and 
gives the axial anomaly 
\cite{abj}
\begin{equation}
\label{anom}
\int\limits_{4m^2}^\infty ds \,\Delta^{(0)}_{\rm pQCD}(s,Q^2) = \frac{1}{2 \pi}.
\end{equation}
According to the Adler-Bardeen theorem \cite{ab}, radiative corrections to the anomaly vanish: Higher order 
QCD calculation can change $\Delta^{(0)}_{\rm pQCD}$ but not the integral.

Non-perturbative QCD effects strongly distort $\Delta(s,Q^2)$ compared with $\Delta^{(0)}_{\rm pQCD}(s,Q^2)$ 
in the low-$s$ region: A meson pole and the hadron continuum are generated. Nevertheless, the integral of 
the entire absorptive part $\Delta(s,Q^2)$ remains unchanged, still representing the anomaly:
\begin{equation}
\label{Aps1}
\int\limits_{0}^\infty ds \,\Delta(s,Q^2) = \frac{1}{2 \pi}. 
\end{equation}
For the case of the isovector $\frac{\bar u u-\bar dd}{\sqrt2}$ axial  
current the spectrum contains the $\pi^0$-meson pole. The physical absorptive part of $F(p^2,Q^2)$ reads 
\begin{eqnarray}
\label{Aps2}
&&\Delta(s,Q^2) =
\\ \nonumber&& 
\pi  \delta (s - m_{\pi}^2) ~ \sqrt{2} f_{\pi}~F_{\pi
\gamma}(Q^2) + \theta (s - s_{\rm th})~\Delta^{I=1}_{\rm cont}(s,Q^2).
\end{eqnarray}
Here $\Delta^{I=1}_{\rm cont}(s,Q^2)$ denotes the hadron-continuum contribution in the isovector channel. 
In (\ref{Aps2}), the $\pi \gamma$ form factor we are interested in appears together with the $\pi$ meson 
$\delta$-function and the pion decay constant.
The anomaly sum rule for $F_{\pi\gamma}(Q^2)$ then takes the form 
\begin{equation}
\label{pigamma}
F_{\pi\gamma}(Q^2) = \frac{1}{2 \sqrt{2}~ \pi^2 f_{\pi}}
\left[1- 2 \pi\int\limits_{s_{\rm th}}^\infty ds\;\Delta^{I=1}_{\rm cont}(s,Q^2)\right].
\end{equation}
For the $\eta\gamma$ and $\eta'\gamma$ form factors, one has to consider the isoscalar currents 
$\bar q q=(\bar uu+\bar dd)/\sqrt{2}$ and $\bar s s$, separately. 
The formulae for $F_{\bar qq}(Q^2)$ and $F_{\bar ss}(Q^2)$ are identical to (\ref{pigamma}) 
except for the replacements $\sqrt{2}f_{\pi}$ by $f_q$ and $f_s$ and $\Delta^{I=1}_{\rm cont}$ by 
$\Delta^{I=0}_{\rm cont}$ and $\Delta^{\bar s s}_{\rm cont}$, respectively. 

For each channel, the relevant threshold $s_{\rm th}$ should be used.
Taking $\eta-\eta'$ mixing \cite{anisovich,feldmann} into account leads to 
\begin{eqnarray}
\label{etaeta}
F_{\eta \gamma} (Q^2)=\frac{5}{3 \sqrt{2}}F_{\bar q q}(Q^2)~
\cos{ \phi} - \frac{1}{3}F_{\bar s s}(Q^2)~\sin{\phi}, 
\nonumber\\
\quad
F_{\eta'\gamma} (Q^2)=\frac{5}{3 \sqrt{2}}F_{\bar q q}(Q^2)~  
\sin{\phi} + \frac{1}{3}F_{\bar s s}(Q^2)~\cos{\phi}. 
\end{eqnarray}
The $\eta-\eta'$ mixing angle $\phi$ is known to be  $\phi \simeq 39^o$; the decay constants are taken to
be $f_q= 1.07 f_{\pi}~,~f_s = 1.36 f_{\pi}$ \cite{feldmann}. 

According to (\ref{pigamma}) and (\ref{etaeta}), the calculation of the  
$P\gamma$ form factors requires an Ansatz for the continuum spectral densities $\Delta_{\rm cont}(s,Q^2)$ 
for all three cases. 

The quark-hadron duality suggests that at large values of $s$, above the resonance region, 
the hadron spectral density should be very close to the perturbative QCD spectral density. 
We therefore use the simple Ansatz 
\begin{eqnarray}
\label{Ct}
&&\Delta_{\rm cont}(s,Q^2)=\theta(s - s_{\rm th})R(s)\Delta^{(0)}_{\rm QCD}(s,Q^2), 
\nonumber\\
&& 
\quad R(s\to\infty)\to 1.  
\end{eqnarray}
It turns out \cite{ms2012} that for the large-$Q^2$ behavior of the form factor the 
behaviour of $R(s)$ at large $s$ is essential: e.g. in order to have the logarithmic rise 
of $Q^2 F(Q^2)$, $R(s)$ should contain a $1/s$-correction: 
For $R(s)=1-r/s$ starting at a finite energy $s_0$, $s_0 > s_{\rm th}$, one finds
\begin{equation}
\label{FF}
Q^2 F( Q^2) \sim \frac{Q^2}{Q^2+s_{0}} (s_{0}-r) + r\log\left(\frac{Q^2+s_{0}}{s_{0}}\right).
\end{equation}
Notice that the part of the integral (\ref{pigamma}) from $s_{\rm th}$ to $s_0$ scales as $1/Q^2$ 
and is therefore not relevant for the effect discussed here.
Thus, neither the details of $R(s)$ at small $s$ nor the presence of higher powers of $1/s$ affect this behavior. 
The negative linear term $-r/s$ in $R(s)$ is responsible for the logarithmic increase. 
If absent, the form factor scales as $F(Q^2)\sim 1/Q^2$. Evidently, for a universal $r$ the high-$Q^2$ 
behavior of all three form factors turns out to be the same. 
For small $r$, the differences due to different thresholds and decay constants affect the 
magnitude of the form factors but not their slope.

\vspace{.6cm}
\section{Conclusions}
We revisited the $P\gamma$ transition form factors, $P=\pi,\eta,\eta'$, with special emphasis on the 
new data on the $\pi\gamma$ form factor reported by Belle \cite{belle2012}. 
Use is made of the exact anomaly sum rule which relates the integral over the hadron spectrum to the axial anomaly. 
This approach has the advantage that the QCD factorization theorem and the meson distribution amplitudes do not 
enter the analyses. Thus, the three processes can be easily compared with each other. 

\noindent $\bullet$ We report that the $\gamma \gamma^*\to P$ form factors of $\pi,\eta$ and $\eta'$ are 
fully compatible with each other---if for the $\pi\gamma$ form factor the recent Belle data are applied: 
the parameter $r$ and the effective threshold used in the description for the $I=0$ nonstrange continuum 
and the ones for the $I=1$ continuum agree with each other.
Thus, the Belle data resolve the puzzle of a qualitatively different behavior 
of the nonstrange component in $\eta$ and $\eta'$ on one hand and of $\pi$ on the other hand.

\noindent $\bullet$ 
The Belle data for the $\pi\gamma$ form factor are compatible with the asymptotic pQCD formula 
indicating that corrections to the asymptotic behavior are small already at $Q^2\ge 10$ GeV$^2$. 

\noindent $\bullet$ 
Still, our {\it best} fits to the data---for all three processes---suggest a slight increase of the product 
$Q^2 F(Q^2)$ at high $Q^2$. If confirmed by future experiments, this would put QCD factorization 
into question and would suggest that the full spectral density of the dispersion representation for the form factor 
is dual to the lowest order pQCD spectral density only by including an effective $1/s$-correction term.

\vspace{.2cm}
\noindent{\it Acknowledgments.} We are grateful to W.~Lucha, B.~Moussallam, J.~Pawlowski, H.~Sazdjian, and  
O.~Teryaev for valuable discussions. D.~M.\ was supported by the Austrian Science Fund (FWF) under Project  
No.~P22843. 




\begin{thebibliography}{30}
\bibitem{cello}
CELLO Collaboration, H.~J.~Behrend et al., Z.~Phys.~C{\bf 49}, 401  
(1991).
\bibitem{cleo}
CLEO Collaboration, J.~Gronberg et al., Phys.~Rev.~D{\bf 57}, 33 (1998).
\bibitem{babar2}
{\sc BaBar} Collaboration, B.~Aubert et al., Phys.~Rev.~D{\bf 74},  
012002 (2006).
\bibitem{babar}
{\sc BaBar} Collaboration, B.~Aubert et al., Phys.~Rev.~D{\bf 80},  
052002 (2009).
\bibitem{babar1}
{\sc BaBar} Collaboration, P.~del~Amo~Sanchez, Phys.~Rev.~D{\bf 84}, 052001 (2011).
\bibitem{radyushkin}
A.~V.~Radyushkin, Phys.~Rev.~D{\bf 80}, 094009 (2009).
\bibitem{roberts}
H.~L.~L.~Roberts, C.~D.~Roberts, A.~Bashir, L.~X.~Gutierrez-Guerrero, P.~C.~Tandy, 
Phys.~Rev.~C{\bf 82}, 065202 (2010).
\bibitem{dorokhov}
A.~Dorokhov, JETP Lett.~{\bf 91}, 163 (2010).
\bibitem{agaev}
S.~S.~Agaev, V.~M.~Braun, N.~Offen, and F.~A.~Porkert, 
Phys.~Rev.~D{\bf 83}, 054020 (2011); arXiv:1206.3968 [hep-ph]. 
\bibitem{teryaev2}
Y.~N.~Klopot, A.~G.~Oganesian, and O.~V.~Teryaev, 
Phys.~Lett.~B{\bf 695}, 130 (2011); Phys.~Rev.~D{\bf 84}, 051901 (2011).
\bibitem{bt}
S.~J.~Brodsky, F.-G.~Cao, and G.~F.~de~Teramond, 
Phys.~Rev.~D{\bf 84}, 033001 (2011); 
Phys.~Rev.~D{\bf 84}, 075012 (2011).
\bibitem{kroll}
P.~Kroll, Eur.~Phys.~J.~C{\bf 71} 1623 (2011).
\bibitem{mikhailov}
A.~P.~Bakulev, S.~V.~Mikhailov, A.~V.~Pimikov, and N.~G.~Stefanis, 
Phys.~Rev.~D{\bf 84}, 034014 (2011); 
Phys.~Rev.~D{\bf 86}, 031501 (2012). 
\bibitem{blm}
I.~Balakireva, W.~Lucha, and D.~Melikhov, 
Phys.~Rev.~D{\bf 85}, 036006 (2012); J.~Phys.~G{\bf 39}, 055007 (2012) [arXiv:1103.3781]; 
W.~Lucha and D.~Melikhov, 
J.~Phys.~G{\bf 39}, 045003 (2012) [arXiv:1110.2080];  
Phys.~Rev.~D{\bf 86}, 016001 (2012) [arXiv:1205.4587]. 

\bibitem{czyz}
H.~Czyz, S.~Ivashyn, A.~Korchin, and O.~Shekhovtsova, Phys.~Rev.~D{\bf 85} 094010 (2012).
\bibitem{lcqm}
C.-C.~Lih and C.-Q.~Geng, Phys.~Rev.~C{\bf 85}, 018201 (2012).
\bibitem{ms2012}
D.~Melikhov and B.~Stech, Phys.~Rev.~D{\bf 85}, 051901 (2012).

\bibitem{belle2012}
Belle Collaboration, S.~Uehara {\em et al.}, arXiv:1205.3249 [hep-ex]

\bibitem{bl}
G.~P.~Lepage and S.~J.~Brodsky, Phys.~Rev.~D{\bf 22}, 2157 (1980).
\bibitem{teryaev}
J.~Horejsi and O.~Teryaev, Z.~Phys.~C {\bf 65}, 691 (1995).
\bibitem{ms}
D.~Melikhov and B.~Stech, Phys.~Rev.~Lett.~{\bf 88}, 151601 (2002).
\bibitem{m}
D.~Melikhov, Phys.~Lett.~B {\bf 380}, 363 (1996); Eur.~Phys.~J.~direct~C {\bf 4}, 2 (2002) [hep-ph/0110087].
\bibitem{abj}
S.~Adler, Phys.~Rev.~{\bf 177}, 2426 (1969); J.~S.~Bell and R.~Jackiw, Nuovo Cimento {\bf 60A}, 47 (1969).
\bibitem{ab}
S.~Adler and B.~Bardeen, Phys.~Rev.~{\bf 182}, 1517 (1969).
\bibitem{anisovich}
V.~V.~Anisovich, D.~I.~Melikhov, and V.~A.~Nikonov, Phys.~Rev.~D{\bf 55}, 2918 (1997); 
V.~V.~Anisovich, D.~V.~Bugg, D.~I.~Melikhov, V.~A.~Nikonov, Phys.~Lett.~B {\bf 404}, 166 (1997).  
\bibitem{feldmann}
T.~Feldmann, P.~Kroll, and B.~Stech, Phys.~Rev.~D{\bf 58}, 114006 (1998); Phys.~Lett.~B{\bf 449}, 339 (1999).
\end{thebibliography}
\end{document}